\documentclass[12pt, a4paper]{article}

\pdfoutput=1

\usepackage{amsmath}
\usepackage{amsfonts}
\usepackage{amssymb}
\usepackage{cite}
\usepackage{multicol,multirow}
\usepackage{graphicx, physics, subcaption}
\usepackage{graphics, epsfig}
\usepackage{epsf}
\usepackage{epstopdf}
\usepackage[usenames,dvipsnames]{color}

\usepackage{colortbl}
\definecolor{summersky}{cmyk}{0.71,0.33,0,0.14}
\definecolor{flamingo}{cmyk}{0,0.51,0.71,0.14}
\definecolor{rp}{cmyk}{0.2, 1, 0.6, 0}
\definecolor{pacificblue}{cmyk}{0.95,0.3,0, 0.19}
\definecolor{gray60}{cmyk}{0.4,0.4,0,0.8}

\numberwithin{equation}{section}

\usepackage[colorlinks,bookmarks=false, hypertexnames=true]{hyperref}
\hypersetup{
	breaklinks=true,
	pdfstartview={FitH},    
	colorlinks=true,       
	linkcolor=blue,          
	citecolor=red,        
	filecolor=magenta,      
	urlcolor=red,           
	anchorcolor=green,      
	linktocpage=true
}

\newcommand{\nc}{\newcommand}

\nc{\ba}{\begin{eqnarray}}
\nc{\ea}{\end{eqnarray}}

\nc{\calR}{{\cal{R}}}
\nc{\calP}{{\cal{P}}}
\nc{\cN}{ {\cal{N}} }

\def\bfk{{\bf k}}

\def\Ylm{Y_{\ell m}(\theta, \phi) }

\def\zlm{Z_{\ell m}}

\begin{document}

\def\thefootnote{\fnsymbol{footnote}}

\begin{center}

{\bf  Quantum Fluctuations in the Interior of Black Holes and Backreactions
}
\\[0.5cm]

{
Hassan Firouzjahi\footnote{firouz@ipm.ir}
}
\\[0.5cm]
 
 {\small \textit{School of Astronomy, Institute for Research in Fundamental Sciences (IPM) \\ P.~O.~Box 19395-5531, Tehran, Iran
}}\\

\end{center}

\vspace{.3cm}
\hrule
\begin{abstract}
We study the propagation of the quantum field perturbations  in the interior of the Schwarzschild black hole. The interior of the black hole is like an anisotropic cosmological background which expands in one extended direction while contracting in azimuthal directions. Solving the quantum mode functions approximately,  we calculate the expectation values of physical quantities such as the energy density $\langle \rho \rangle$ and pressure $\langle P\rangle$ as measured by an observer in the interior of the black hole. We employ a combination of the $\zeta$ function regularization and dimensional regularization schemes to regularize the UV divergences in the energy momentum tensor associated to these quantum perturbations.  
By solving the Einstein field equations, we calculate  the quantum 
backreactions induced in the background geometry. We speculate that the effects of quantum fluctuations  in the interior of the black hole can be measured by  the exterior observer. 
\end{abstract}

\vspace{0.3cm}
\hrule

\newpage

\section{Introduction}
\label{sec:intro}

The event horizon surrounding a black hole (BH) separates it into two causally disconnected regions. To a remote observer outside the BH, it represents a point-like singularity which may be observed via their gravitational effects such as in merging and generation of  gravitational wave  as observed in LIGO/Virgo/Advanced Virgo collaborations \cite{LIGOScientific:2016aoc, LIGOScientific:2016sjg}.  Most of the studies concerning BHs are naturally devoted  to the exterior region as viewed by the exterior observers. This is mainly motivated from the simple fact that the events occurring inside the BH are inaccessible to the exterior observer. Having said this, the interior of the BH reveals interesting physical phenomena which deserve more attention. Specifically, the interior of BH looks like a cosmological background in which the singularity is spacelike  in  the form of big crunch singularity.  In simple setups such as in Schwarzschild solution, the interior of BH looks like a homogenous but  anisotropic cosmological background. The  spatial part has the topology of  $R\times S^2$ known as the Kantowski-Sachs  spacetime \cite{Kantowski:1966te}. This anisotropic cosmological background contains two scale factors, and correspondingly, two ``Hubble" expansion rates. The scale factor along the extended spatial direction expands while the scale factor associated to the azimuthal directions contracts. Eventually, near the singularity, the scale factor  of the azimuthal direction reaches zero, signalling the 
onset of the big crunch singularity. 

In the past there have been works to treat the interior of a BH as a cosmological background. One interesting idea is  that   the  interior of BH is  replaced by a non-singular dS spacetime, as  studied in   \cite{Sakharov:1966aja, Poisson:1988wc, Frolov:1989pf, Frolov:1988vj, Easson:2001qf, Firouzjahi:2016nle, wenda1988junction, Yoshida:2018kwy, Brandenberger:2021ken, deCesare:2020swb, Gaztanaga:2021rnv, Gaztanaga:2022gbd} or the proposal of 
gravastars  \cite{Mazur:2001fv, Visser:2003ge, Cattoen:2005he}.   
The main motivation  in these studies was  to remove  the singularity of BH. On the physical grounds one expects that the singularity of BH is a shortcoming of the classical  general relativity. It is expected that on very small scales, say the Planck scale, the quantum gravity effects play important roles to resolve the classical singularity. In particular, 
the proposal of maximum curvature of spacetime  
\cite{markov1982limiting, Frolov:1989pf, Frolov:1988vj} is an interesting idea in this direction.

The question of quantum field perturbations in the exterior of the BH is vastly studied. Among the non-trivial issues in these studies is the fact that there is no unique vacuum in a curved background, specially in a BH background which has an event horizon. This has lead to the celebrated  Hawking radiation effect \cite{Hawking:1975vcx, Unruh:1976db}, for some reviews of Hawking radiation see \cite{Mukhanov:2007zz, Townsend:1997ku, Fabbri:2005mw, Visser:2001kq, Jacobson:2003vx}.  While QFTs effects are vastly studied in the exterior region, they are less investigated for the interior region.
In this work we study the propagation of quantum field perturbations in the interior of BH. Intuitively speaking, the perturbations enter the horizon after crossing the event horizon and propagate towards the singularity.  Since the interior of the BH is like an anisotropic cosmological background, the propagation of the quantum perturbations in the interior region is similar to the evolution  of perturbations in FLRW cosmology. While this 
similarity is very helpful for the understanding of the nature of quantum perturbations in the interior of BH, but there are a number of important differences as well. First, compared to isotropic FLRW cosmology, we have less symmetry, the former enjoys an $O(3)$ symmetry while the latter has the
smaller $O(2)$ symmetry. The other important difference is that in FLRW cosmology the quantum perturbations are generated near the time of big bang, as in inflationary cosmology. Here, however, the perturbations are viewed to be generated near the event horizon and terminated at the moment of big crunch.   
 
In this study we consider a massless scalar field in the interior of a Schwarzschild background. After reviewing the background cosmology 
associated to the interior of the BH (section \ref{prelim}), we consider the quantum perturbations. Upon decomposing the perturbations into appropriate bases of the creation and annihilation operators, we solve the mode function approximately (section \ref{QFT}).  Our main interest is to calculate the expectation values of physical quantities such as the energy density $\langle \rho \rangle$ and pressure $\langle P\rangle$. As in many QFT analysis, we inevitably encounter  UV divergences which should be regularized 
appropriately (section \ref{regulrization}). Our goal is to calculate the backreactions of the regularized energy momentum tensor associated to these quantum perturbations on the background geometry. Upon solving the Einstein field equations, we obtain the corrections in the metric to leading order in perturbations and look for their physical implications (section \ref{backreaction}).

 \section{Background Geometry}
\label{prelim}

In this section we briefly review basic properties of the Schwarzschild solution 
which will be employed for our treatment of the quantum fields in this background.

\subsection{Black Hole Preliminaries}

We consider the Schwarzschild background with the following metric, 
\ba
\label{metric-Sch}
\dd s^2 = - \big(
1- \frac{2GM}{r}\big) \dd {\bar t\, }^2 + \frac{\dd r^2}{\big(
	1- \frac{2GM}{r}\big)} + r^2 \dd \Omega^2 \, ,
\ea
in which $G$ is the Newton constant, $M$ is the mass of the BH as measured by a distant  observer and  $\dd\Omega^2$ represents the metric on a unit two-sphere. The coordinate system $(r,\bar t)$  covers only the exterior of the whole manifold which becomes singular on the BH event horizon at $r= R_S\equiv 2 GM$. Note that we use the notation $\bar t$ for the time coordinate employed by the exterior observer while we reserve $t$ for a timelike coordinate employed for the interior of the black hole. 
Naively speaking, for the interior of the BH the roles of $\bar t$ and $r$ coordinates are switched in which $\bar t$ becomes spacelike while $r$ becomes timelike. This also suggests that the interior of BH is dynamical, as in cosmological backgrounds. 

Since the coordinate $(r,\bar t)$ is singular at the horizon,  to cover the entire manifold we can use the  Kruskal-Szekeres coordinate $(T, R)$ which are 
related to the original  coordinate $(r,\bar t)$ by
\ba
\label{UV-r}
T^2 - R^2 =  e^{r/2GM} \big(
1- \frac{r}{2GM}
\big)  \, ,
\ea
and
\ba
\label{UV-t}
\frac{T}{R}  = \tanh(\frac{\bar t}{4 G M}) \, .
\ea
Furthermore, it is very convenient  to use the  Kruskal-Szekeres coordinate in its lightcone base $(U, V)$ defined via 
\ba
U\equiv GM( T-R),  \quad  \quad V\equiv GM( T+R) \, .
\ea
We use the convention such that  the coordinates $(U, V)$ carry the dimension of length (or time). 

\begin{figure}[t]
\vspace{-3cm}
\begin{center}
	\includegraphics[scale=0.4]{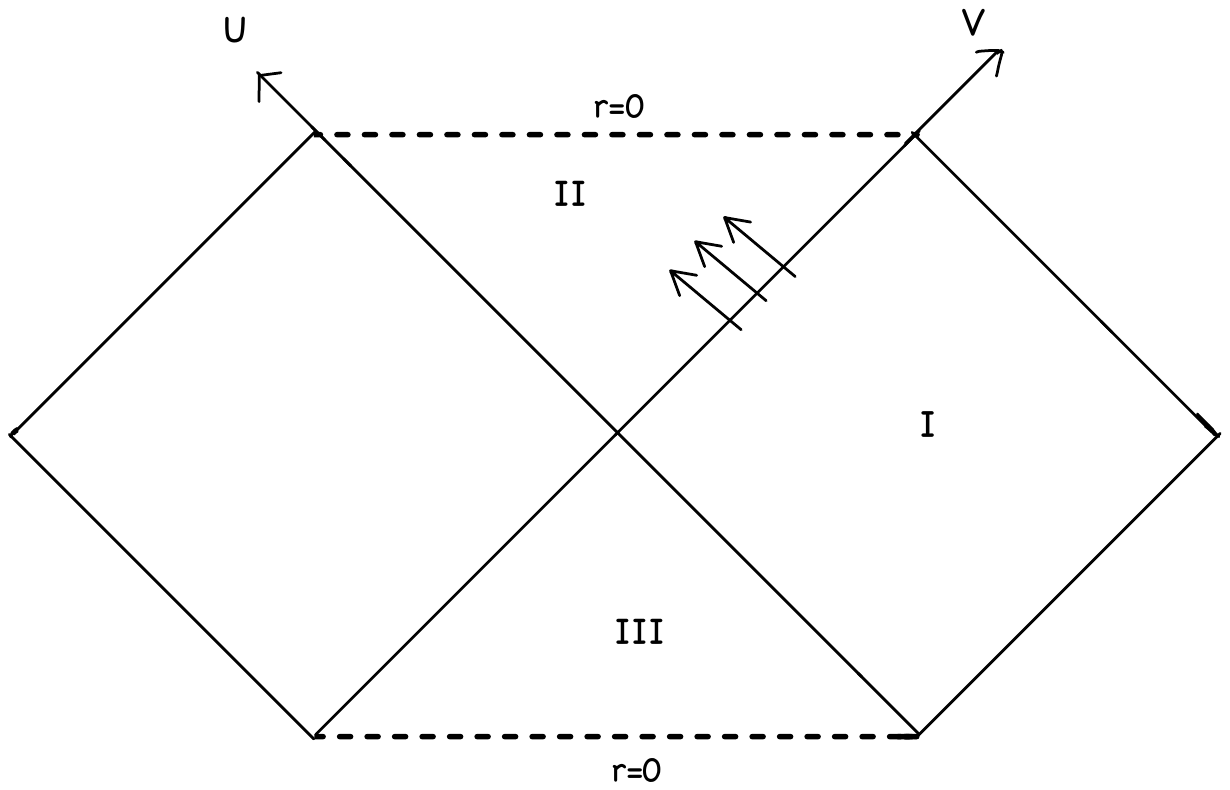}
	\end{center}
	\vspace{-3cm}
\caption{ The Kruskal-Szekeres diagram. Region I (II) denotes the exterior (interior) of BH while the region III denotes the WH part. The future event horizon is located at $U=0, V>0$. The arrowed lines represent the propagation of the quantum perturbations towards the interior region, heading to future singularity (big crunch) at $r=0$. 
\label{Kruskal}
}
\vspace{0.5cm}
\end{figure}

A conformal diagram of the entire Schwarzschild manifold is depicted in Fig. \ref{Kruskal}. The exterior of the BH is the region $V>0, U <0$ while the interior of the BH, which we are interested in here,  is the region $U, V>0$ bounded by the  future singularity  $r=0$. The white hole (WH)  region is given by $U, V<0$ bounded by the past singularity $r=0$. Both the interiors of the BH and the WH share the common property that their backgrounds are dynamical,  corresponding to anisotropic cosmological setups. However, the crucial difference is that the singularity of the BH is in future, i.e. it represents a big crunch singularity while the singularity of the WH is in the past, i.e. it represents a big bang singularity. Mathematically speaking, the WH solution 
is realized from the time reversal symmetry of the BH solution in classical general relativity. 

The generation and propagation of the quantum perturbations in WH region were studied in \cite{Firouzjahi:2022rtn} as a realization of Hawking radiation measured by an observer far outside the BH. More specifically, it was argued in \cite{Firouzjahi:2022rtn} that the quantum fluctuations which are generated at past singularity $r=0$ (big bang) will propagate towards the past horizon $V=0, U<0$ and eventually towards ${\cal I}^+$ which is measured by an observer at future infinity in the exterior part of the BH. It was shown that while the observer deep inside the WH and the observer far outside the BH both share the same vacuum, but their vacuum  is different than the one defined for a freely falling observer near the horizon. In particular, the WH observer and the observer far outside the BH detect a spectrum of particle with the Planck distribution having  the Hawking temperature  $T_H = \frac{1}{8 \pi GM}$. Intuitively speaking, the conclusion of \cite{Firouzjahi:2022rtn}  implies that  if a BH is not entirely black due to Hawking radiation, then the WH is not entirely white either!

In this works, however, we consider the quantum fluctuations which enters inside the BH from the future event horizon $U=0, V>0$ and propagate towards the future singularity $r=0$ (big crunch) as shown schematically by the arrowed lines in  Fig.  \ref{Kruskal}.



\subsection{Cosmology of the Interior of the Black Hole }
\label{BH-cosmology}

As mentioned above, the interior of the BH  represents an anisotropic cosmological setup known as the Kantowski-Sachs background \cite{Kantowski:1966te, Doran:2006dq}  with the spatial part having the structure of  $R \times S^2$.  To see this more clearly, let us 
start with the usual  tortoise coordinate 
\ba
\label{r*r-eq}
\dd r_* = (1- \frac{2GM}{r})^{-1} \dd r \, ,
\ea
which for the interior of BH  ($r<2GM$) yields the following relation, 
\ba
\label{r*-WH}
r_* = r + 2 GM \ln \left( 1- \frac{r}{2GM} \right) ,   \qquad \quad ({r < 2 GM}) \, .
\ea
We note that  $-\infty < r_* \leq 0$ in which $r_*=-\infty$ corresponds to the future event horizon $r=2GM, U=0$ with $V>0$
while $r_* =0$ corresponds to the future singularity at $r=0$.  

Let us define the future directed time coordinate $\dd\tau=\dd r_*$ so
\ba
\label{tau-def}
\tau = r + 2 GM \ln \left( 1- \frac{r}{2GM} \right) \qquad \quad ({r < 2 GM}) \, .
\ea
Correspondingly, the BH singularity is at $\tau=0$ while the future event horizon  $V>0, U=0 $ is mapped to $\tau=-\infty$. Proceeding further, let us  define the spacelike coordinate  $\dd x \equiv \dd \bar t$ with $-\infty < x < +\infty$. In the new coordinate $(\tau, x)$,  the metric for the interior of BH  takes the following cosmological form
\ba
\label{BH-metric}
\dd s^2 = a(\tau)^2 ( -\dd \tau^2 + \dd x^2) + r(\tau)^2 \dd \Omega^2 \, ,
\ea
in which the scale factor $a(\tau)$ is defined via
\ba
\label{a-scale}
a(\tau) \equiv  \Big( \frac{2 G M}{r(\tau)} -1  \Big)^{\frac{1}{2}} \, . 
\ea
Note that from Eq. (\ref{tau-def}), one can solve for $r$ as a function 
of $\tau$. 

The metric (\ref{BH-metric}) represents an anisotropic cosmological background
with two scale factors $a(\tau)$ and $r (\tau)$. Near the horizon, $a(\tau) \rightarrow 0$ while $r(\tau) \rightarrow R_S$. As time goes by, the space along the $x$ direction expands  while the scale factor along the two-sphere shrinks. At the moment of big crunch $\tau=0$, $a(\tau) \rightarrow \infty$ while $r(\tau) \rightarrow 0$. 

It is insightful to look at the metric for two limiting times,  $\tau \rightarrow -\infty$ and  $\tau \rightarrow 0$ more closely. Using Eq. (\ref{tau-def}), at later time when  $r \rightarrow 0$  we have 
\ba 
\label{ra-app}
r^2 \simeq - 4 G M \tau \, ,  \quad \quad  a(\tau) \simeq \big( \frac{G M}{-\tau}\big)^{\frac{1}{4}} \quad \quad \quad  (r \rightarrow 0) \, ,
\ea
so the metric takes the following form
\ba
\label{limit1}
\dd s^2 \simeq  \sqrt{\frac{GM}{-\tau}} ( - \dd\tau^2 + \dd x^2) - 4 G M \tau ~\dd \Omega^2
\quad \quad (\tau \rightarrow 0^-) \, .
\ea
The contraction along the $x$ direction and the expansion along the azimuthal directions is evident from the above line element.

On the other hand, near the past horizon $r=2GM, U=0 $,  we have, 
\ba
r(\tau) \simeq 2 GM \Big[ 1- \exp \big(\frac{\tau}{2 GM} \big)
\Big] \quad \quad (\tau \rightarrow -\infty) \, ,
\ea
so the metric takes the following form,
\ba
\label{limit2}
\dd s^2 \simeq \exp \big({\frac{\tau}{2 G M}} \big) ( - \dd\tau^2 + \dd x^2) + 4 G^2 M^2  \dd \Omega^2
\quad \quad (\tau \rightarrow -\infty) \, .
\ea
Note that near the horizon, the two-sphere reaches the fixed radius $r \rightarrow  R_S$ while the scale factor in the $x$-direction falls off exponentially.  Although the scale factor 
$a(\tau)$ falls off, but the spacetime is regular near the horizon. This is because the original coordinate $(r, \bar t)$ was singular near the horizon. More specifically, calculating the Riemann tensor associated to the metric (\ref{limit2}) we find that $R_{\theta \phi \theta \phi} = 4 G^2 M^2 \sin(\theta)^2$ while all other components are zero so the background is regular near the horizon as expected. 

We can define the ``cosmic time" $\hat t$ related to the conformal time via $\dd \hat t \equiv a(\tau) ~\dd \tau$ so 
\ba
\label{hat t-r-eq}
\dd \hat t = \frac{-\dd r}{ \sqrt{\frac{2 G M}{r} -1}} \, .
\ea
Choosing the onset of big crunch singularity to be at $\hat t=0$, we obtain
\ba
\hat t=  \sqrt{r ( 2 G M -r)} + M G  \Big[-\frac{\pi}{2}+ \mathrm{arc} \sin \Big( 1- \frac{r}{G M} \Big)  \Big] \, ,
\ea
so $ -{\pi} GM <    \hat t < 0$ inside the BH.  
From the inverse of the above expression, one can obtain $r= r(\hat t)$  so the metric in term of cosmic time is given by
\ba
\label{metric-cosmic}
\dd s^2 = - \dd \hat t^2 + \Big( \frac{2 G M}{r (\hat t\, )} - 1 \Big) \dd x^2 + 
r (\hat t\, )^2 \dd \Omega^2. 
\ea

It is instructive to look at the geometry near the big crunch singularity when expressed in the form of metric (\ref{metric-cosmic}).  Near $r=0$, we have 
\ba
r \simeq ( G M)^\frac{1}{3}  \bigg(\frac{-3 \hat t}{\sqrt2} \bigg)^\frac{2}{3}
\quad \quad (\hat t \rightarrow 0^-) \, ,
\ea
so the metric (\ref{metric-cosmic}) takes the following form, 
\ba
\label{metric-cosmic2}
\dd s^2 \simeq -  \dd \hat t^2 +  ( G M)^\frac{2}{3} \Big( c_1 {\hat t}^{-\frac{2}{3}} \dd x^2 + c_2 {\hat t}^{\frac{4}{3}} \dd \Omega^2 \Big) \, ,
\ea
in which $c_1$ and $c_2$ are two numerical factors.  The above geometry is in the form of  the Kasner  metric. More specifically, the Kasner metric is the solution of the Einstein field equation in the vacuum 
with the following form, 
\ba
\label{Kasner}
\dd s^2 = - {\dd \hat t\, }^2 + \sum_{i=1}^3 {\hat t \,}^{2 p_i} \dd x_i^2 \,  ,
\ea
where the exponents $p_i$ satisfy the following constraints:
\ba
\label{constraints-Kasner}
\sum_{i=1}^3 p_i = \sum_{i=1}^3 p_i^2 =1 \, .
\ea
From the  above  constraints one concludes that  at least one of the 
exponents $p_i$ should be negative.  In our example of BH cosmology near the big crunch singularity with the metric (\ref{metric-cosmic2}), one has $p_1= -\frac{1}{3}$ while  $p_2= p_3 =\frac{2}{3}$. It can be checked that these exponents satisfy the constraints (\ref{constraints-Kasner}) with the contraction being along the 
{$x$} direction.

We can also define the ``Hubble  rate" associated to metric (\ref{metric-cosmic}).  Since there are two scale factors $a(\hat t)$ and $r(\hat t)$, correspondingly we should define two Hubble rates. Defining the Hubble rate associated to the scale factor $a(\hat t)$ by $H_1$, we obtain
\ba
 H_1\equiv \frac{1}{a}  \frac{\dd a}{ \dd \hat t}=  \frac{G M}{{r (\hat t )}^2} \Big( \frac{2 G M}{r(\hat t\, )} -1 \Big)^{-\frac{1}{2}} \, .
\ea 
Since the scale factor $a$ is expanding, the Hubble  rate $H_1$ is positive. 
Furthermore, $H_1$ diverges both near the BH big crunch singularity and near the horizon. The former divergence is genuine as the spacetime is singular at $r=0$ so physical quantities, such as the  energy density or Hubble rate, diverge. On the other hand, the  divergence  of $H_1$ near the  event horizon is an artifact of coordinate singularity as suggested from the relation between $\dd \hat t$ and $\dd r$ in Eq. (\ref{hat t-r-eq}).

On the other hand, denoting the Hubble rate associated to the scale factor $r(\hat t)$ by $H_2$  we have, 
\ba
H_2\equiv \frac{1}{r(\hat t) }  \frac{\dd r(\hat t) }{ \dd \hat t}= \frac{-1}{r(\hat t)} \Big( \frac{2 G M}{r(\hat t \, )} -1 \Big)^{\frac{1}{2}} \, .
\ea
Since the scale factor  $r(\hat t)$ is contracting,  $H_2 <0$. In addition,  $H_2$ diverges near the singularity as expected but it vanishes near the event horizon. This is expected since near the horizon ($\tau \rightarrow -\infty$) the radius of the  two-sphere approaches a  constant size so $H_2$ vanishes in this limit. From the  expressions for $H_1$ and $H_2$, one can check that the following relation holds,
\ba
H_1 H_2 = -\frac{G M}{r (\hat t\, )^3} \, ,
\ea
so the two Hubble rates have opposite signs as required from the structure of Kasner geometry (i.e. expansion in one direction and contractions in the remaining directions).


\section{Quantum Fields Inside the Black Hole }
\label{QFT}

Here we study the quantum field perturbations  in the interior of BHs.
A similar analysis were performed for the quantum fluctuations in the WH backgrounds in  \cite{Firouzjahi:2022rtn}, see also \cite{Gusin:2023abp}
for a recent study of the quantum field in the interior of the BH.  

We consider a real massless scalar field $\Phi$ which propagates in the interior of the BH.   In this section, we work in the test field limit where the backreaction of the field on the background is negligible. However, in the next section we take into account the effects of the backreactions on the background geometry.  We use the  metric (\ref{BH-metric}) with the conformal time $\tau \equiv r_*$ while $x \equiv \bar t$ represents the extended spatial 
direction (as mentioned before,  in the interior of the BH, $\bar t$  is spacelike while 
$r$ is timelike). 

To perform the QFT regularizations, we employ the dimensional regularization scheme \cite{tHooft:1972tcz, tHooft:1974toh, Bollini:1972ui, Deser:1974cz, Dowker:1975tf, Barvinsky:1985an} in which the extended directions ($R^2$ parts of the whole $R^2 \times S^2$ manifold) is generalized to $d$ dimensions so the whole manifold now is $R^d \times S^2$. The $R^d$ part has one timelike coordinate along $\tau$ as before while the $d-1$ spatial directions are denoted by $\bf{x}$.  
 To perform the regularization, we take $d=2 -\epsilon$ and take $\epsilon \rightarrow 0$ at the end. The divergent terms with the poles like $\epsilon^{-1}$ in physical quantities such as the energy density or pressure 
are discarded as these UV divergences are taken care of by appropriate counter terms. The regularized contributions are the finite $\epsilon^0$ terms.  
 
With these discussions in mind, the action of the scalar field in $d+2$ dimension is given by
\ba
\label{action}
 S &=&-\frac{1}{2} \int \dd \tau \dd  x^{d-1} \dd \Omega\,  a(\tau)^{d} r(\tau)^2  \partial^\mu \Phi \partial_\mu \Phi \, \nonumber\\
 &=& \frac{1}{2}  \int \dd \tau \dd  x^{d-1} \dd \Omega\, a(\tau)^{d-2} r(\tau)^2 
 \Big[ \big( \partial_\tau \Phi \big)^2 - \big( \partial_{\bf x} \Phi \big)^2-
 \frac{a(\tau)^2}{r(\tau)^2} \gamma^{a b} \partial_a \Phi \partial_b \Phi  \Big]\, ,
 \ea
 in which $\gamma_{ab}= \mathrm{diag} (1, \sin^2 \theta)$ is the metric along the azimuthal directions $\{ a, b\}= \{ \theta, \phi\}$. 
Note that the factor $a(\tau)^{d}$ in the first line appears because the part of the background metric  (\ref{BH-metric}) endowed with the scale factor $a(\tau)$
is $d$-dimensional.

Since the background enjoys the rotation symmetry in the azimuthal directions,  we  expand the quantum field $\Phi$ in terms of the spherical harmonics  $Y_{\ell m} (\theta, \phi)$ as follows, 
\ba
\label{Phi-Z}
\Phi = f(\tau)  \sum_{\ell =0}^\infty \sum_{ m=-\ell}^\ell   Z_{\ell m}(\tau, {\bf x}) Y_{\ell m} (\theta, \phi)  \, ,  \quad \quad f(\tau) \equiv \frac{a(\tau)^{\frac{2-d}{2}}}{r(\tau)} , 
\ea
in which $Z_{\ell m}(\tau, {\bf x})$ should be viewed as the  quantum mode operator. We have pulled out a factor $f(\tau) = a^{(2-d)/2}/r$ so  $Z_{\ell m}$ will be the canonically normalized field in the perturbation analysis.  Note that when  $d=2$, we recover the usual convention that $f=1/r$ as employed in BH quasi-normal mode analysis.  Finally, the reality of $\Phi$ requires that $Z_{\ell m}^* = (-1)^m Z_{\ell\,  -m}$. 

Plugging the  decomposition (\ref{Phi-Z}) for the quantum field in  the action (\ref{action}) we obtain the following canonical form for the action, 
\ba
\label{action-canonic}
S=\frac{1}{2} \int \dd \tau \dd x^{d-1}  \sum_{\ell, m}
\Big[  \big | \frac{\partial Z_{\ell m}}{\partial \tau} \big|^2
- \big | \frac{\partial Z_{\ell m}}{\partial {\bf x}} \big|^2 +\Big( 
  \frac{2f'^2}{f^2}  - \frac{f''}{f}  + \frac{\ell (\ell +1)}{r^2} \big(1- \frac{2 GM}{r}\big) \Big) 
|Z_{\ell m}|^2
\Big] ,
\ea
in which a prime denotes a derivative with respect to conformal time 
$\tau=r_{*}$.  To obtain the above result, the following relations concerning the spherical harmonics have been used, 
\ba
\int \dd \Omega\,  \Ylm Y_{\ell' m'}^*(\theta, \phi) = \delta_{\ell \ell'} \delta_{m m'} \, , 
\ea
and,  
\ba
\label{sum-Ylm}
Y_{l m}(\theta, \phi)^* = (-1)^m  Y_{l \,  -m}(\theta, \phi) \, ,
\quad \quad 
\sum_{m=-\ell}^{\ell} | Y_{\ell m} (\theta,\phi)|^2 =\frac{2 \ell +1}{4 \pi} \, .
\ea

Since the spatial coordinate $-\infty < {\bf x} < \infty$ has the  translation invariance we can Fourier expand  $\zlm \propto e^{-i \bfk \cdot {\bf x}}$. Correspondingly, the quantum mode $Z_{\ell m}(\tau, x)$   is expanded in 
$(d-1)$-dimensional Fourier space as follows,
\ba
\label{mode-Z-tor}
Z_{\ell m} (\tau, {\bf x}) = \int_{-\infty}^{+\infty} \frac{\dd^{d-1} \bfk}{(2 \pi)^{(d-1)/2}} 
\left[  e^{i \bfk \cdot {\bf x}} g_k^\ell(\tau) b_{\bfk}^{\ell m} 
+ (-1)^m   e^{-i \bfk \cdot {\bf x}}{g_k^\ell}^*(\tau) 
{b_{\bfk}^{\ell\,  -m}}^\dagger   \right] \, .
\ea
Here $b_\bfk^{\ell m}$ and ${b_{\bfk}^{\ell m}}^\dagger$  are the annihilation and the creation operators defined for the observer inside the BH which 
satisfy the usual commutation relations, 
\ba
[ b_\bfk^{\ell m}, {b_{\bfk'}^{\ell' m'}}^\dagger] = \delta_{\ell \ell'} \delta_{m m'}
\delta (\bfk - \bfk') \, , \quad 
[ b_\bfk^{\ell m},   b_{\bfk'}^{\ell' m'}] = [ {b_{\bfk}^{\ell m}}^\dagger, {b_{\bfk'}^{\ell' m'}}^\dagger] = 0 \, .
\ea
Note that the factor $(-1)^m$ appears in the second term of the mode expansion to enforce the reality of $\Phi$. 

From the action (\ref{action-canonic}) the equation of the mode function in Fourier space is obtained as follows,  
\ba
\label{mode-eq}
\frac{\dd^2 g^\ell_k(\tau) }{\dd \tau^2}
+ \Big( k^2 -  V_{\mathrm{eff}}(\tau) \Big)  g^\ell_k(\tau) =0 \, ,
\ea
in which the effective potential is given by, 
\ba
\label{Veff}
V_{\mathrm{eff}} (\tau) =  \frac{2f'^2}{f^2}  - \frac{f''}{f} 
+ \frac{\ell (\ell +1)}{r^2} \big(1- \frac{2 GM}{r} \big) \, .
\ea
Eq. (\ref{mode-eq}) is similar to the  standard Regge-Wheeler perturbation equation for the field in the exterior region of the BH \cite{Regge:1957td}. However, in our setup where the interior of BH is treated as a cosmological background, the above equation can be interpreted as the  extension of the Sasaki-Mukhanov equation  to the anisotropic cosmological background. Technically speaking,   the mode function in our analysis is time-dependent (i.e. the differential equation is with respect to time $\tau$)  while in the standard Regge-Wheeler equation the relevant coordinate to solve the field equation is spacelike, the coordinate $r_*$ in the exterior of BH. 

With some efforts, and using the relations (\ref{r*r-eq}), one can show that 
\ba
\label{Veff}
V_{\mathrm{eff}} (\tau) = \frac{2 G M}{r(\tau)^3} \Big(1- \frac{(6-d) (2+d) GM}{8 r(\tau)} \Big) 
+ \frac{\ell (\ell +1)}{r(\tau)^2} \big(1-  \frac{2 GM}{r(\tau)}\big)  \, .
\ea 
One can check that for $d=2$, this yields the standard form of 
$V_{\mathrm{eff}}$ as  in BH quasi-normal mode analysis 
\cite{Nollert:1999ji, Kokkotas:1999bd}.

Note that  $V_{\mathrm{eff}} (\tau)<0$. It goes to zero  near the horizon corresponding to $\tau \rightarrow - \infty$ and diverges near the singularity at $\tau \rightarrow 0^-$. We try to solve  Eq. (\ref{mode-eq}) approximately 
as it can not be solved analytically for the entire  region inside BH for $-\infty < \tau <0$. For the regions near the horizon the effective potential  falls off rapidly and the dominant term in the  bracket in  Eq. (\ref{mode-eq}) is simply $k^2$. The mode functions then  are the plane wave solutions $g_k(\tau) \propto e^{\mp i \omega \tau}$ in which $\omega \equiv  |k|\ge   0$.  Since we assume that the perturbations  are generated near the horizon and move forward towards the big crunch singularity (future singularity), we take the positive frequency solution 
$g_k(\tau) \propto e^{- i \omega \tau}$ as the incoming solution. On the other hand,  near the singularity the contribution of the term containing $\ell( \ell+1)$ can be neglected compared to the first term in $V_{\mathrm{eff}} (\tau)$ which is more divergent. As a result, as argued in \cite{Firouzjahi:2022rtn}, we can neglect the contribution of the 
term  $\ell( \ell+1)$ and the term containing $r^{-3}$ in the effective potential 
and consider the most divergent term $r^{-4}$.

Near  the singularity $ r^2  \simeq -4 GM \tau$ so within our approximation,  Eq.  \eqref{mode-eq} simplifies to
\ba
\label{mode-eq1}
g_k''(\tau)  + \left(  k^2 + \frac{(6-d) (2+d)}{64 \tau^2}  \right)  g_k(\tau)=0 \, 
\quad \quad (\tau \rightarrow 0) \, ,
\ea
in which a prime  denotes the derivative with respect to $\tau$. 
To simplify the notation, we have dropped the superscript $\ell$ in $g_k^\ell(\tau)$ so  effectively we work in the $s$-wave limit with  $\ell=m=0$.

As mentioned before, Eq. (\ref{mode-eq1}) is similar to the Sasaki-Mukhanov equation  in  cosmological perturbation theory. Therefore,  interesting insights can be obtained  when compared to perturbations in FLRW background.  We have two scale factors $r(\tau)$ and $a(\tau)$, in which the latter is expanding as time goes by (i.e. as we move forward towards the big crunch singularity). Since we have decomposed the perturbations into spherical harmonics, Eq. (\ref{mode-eq1}) represents the dynamics of perturbations in a 1+1 expanding universe with the  scale factor $a(\tau)$ given in Eq. (\ref{a-scale}). Similar to the studies of perturbations in FLRW cosmology, we
can divide the perturbations into long ``superhorizon" and short ``subhorizon" perturbations.  Perturbations which satisfy $ k |\tau| \gg 1$ are initially inside the horizon. As we move towards big crunch and  $\tau\rightarrow 0$, $ k |\tau| \ll 1$ and the mode becomes superhorizon.  Eventually, at the moment of big crunch,  all modes become superhorizon. 

Fortunately Eq. (\ref{mode-eq1}) can be solved with the solution, 
\ba
\label{gk-sol}
g_k (\tau) = \sqrt{-\tau} \left[ C_1(\omega) H_{\nu}^{(1)}(-\omega \tau) + C_2(\omega)  H_{\nu}^{(2)}(- \omega \tau)  
\right] \, ,  \quad \quad \nu \equiv \frac{d-2}{8} \, ,
\ea
in which  $H_{\nu}^{(1, 2)}$ are the Hankel functions and  $C_1(\omega), C_2(\omega)$ are constants of integrations. As discussed before, the solution \eqref{gk-sol}  is an approximate solution  since we have neglected various subleading terms in the effective potential.  However, if we manage to connect the incoming plane wave solution to the near singularity solution \eqref{gk-sol} then we can be hopeful that the resultant solution mimics the true solution to reasonable accuracy. This is possible because the Hankel functions asymptotically approach to the plane wave solutions  $ e^{\mp i \omega \tau}$. 
This suggests that the solution  (\ref{gk-sol}) is a good approximation for both  regions   $ \tau \rightarrow 0$ and $\tau \rightarrow -\infty$ with some modifications in the intermediate region. With this approximation, we take  the analytic solution (\ref{gk-sol}) to be  qualitatively valid for the entire region $-\infty < \tau <0$. 

Our job is now to fix the constants $C_1(\omega), C_2(\omega)$. This can be achieved by  imposing the equal time quantum commutation relation between the field $Z$ and its conjugate momentum $\partial_\tau Z$, 
\ba
\label{quan.}
[ Z (\tau, x) , \partial_{\tau} Z(\tau, x') ] =  i \delta (x-x') \, .
\ea
Imposing the above quantization condition in our mode function (\ref{gk-sol}) yields the following relation, 
\ba
\label{norm}
\big| C_1(\omega) \big|^2 - \big|C_2(\omega) \big|^2 = \frac{\pi}{4} \, .
\ea

The above constraint does not uniquely fix $C_1(\omega)$ and $C_2(\omega)$. However, the above equation  suggests the natural possibility that  $C_1 = \sqrt \pi/2$ and $C_2=0$. Indeed one can fix 
$C_1$ and $C_2$ definitely only considering the initial conditions near the horizon where the perturbations are generated. The initial condition was fixed by the positive frequency mode with $g_k(\tau) \propto e^{-i \omega \tau}$.
Using the asymptotic form of the Hankel function, 
\ba
\label{Hankel-asym}
H_{\nu}^{(1)}( -\omega \tau) \rightarrow \sqrt{\frac{-2}{ \pi {\omega} \tau}} e^{-i \big(\omega \tau + \frac{\pi}{4}{(1+2 \nu)} \big) } \quad \quad (\tau \rightarrow -\infty) \, ,
\ea
one can check that only the term containing $C_1$ in Eq. \eqref{gk-sol}
can match to the positive frequency mode while the term containing $C_2$ match to the negative frequency mode $e^{i \omega \tau}$. This justifies our conclusion that 
$C_2=0$. This conclusion is similar  to the standard ``Bunch-Davies" vacuum in a dS background  which carries the lowest energy among all possible vacua. Of course, if we consider a mixed initial condition with both positive and negative frequencies, then one can consider the general case in which both $C_1(\omega)$ and $C_2(\omega)$ are present, similar to  the ``non Bunch-Davies" vacuum in inflation.   With $C_2=0$, we finally obtain the mode function as follows 
\ba
\label{g*-rt}
g_k (\tau) = \sqrt{\frac{-\pi \tau}{4}}  H_{\nu}^{(1)}( -\omega \tau)  \, .
\ea

In summary, putting all together, the quantum perturbation 
$\Phi(x^\mu)$ from Eq. (\ref{Phi-Z}) is obtained to be
\ba
\label{Phi-pert}
\Phi(x^\mu) =\frac{a(\tau)^{\frac{2-d}{2}}}{r(\tau)} \sum_{\ell, m}
\int \frac{\dd^{d-1} \bfk}{(2 \pi)^{(d-1)/2}} 
\left[  e^{i \bfk \cdot {\bf x}} g_k(\tau) b_{\bfk}^{\ell m} + (-1)^m   e^{-i \bfk \cdot \bf{x}} g^*_k(\tau) {b_{\bfk}^{\ell\, -m}}^\dagger  \right]  \Ylm  \, ,
\ea
with the mode function $g_k(\tau)$ given in Eq. (\ref{g*-rt}).

\section{The Regularization}
\label{regulrization}

Armed with the mode function given in Eq. (\ref{Phi-pert}), we can calculate the energy momentum tensor associated to quantum perturbations. Technically speaking, these quantum corrections correspond to one-loop correction from vacuum bubble diagrams. Similar studies were performed for dS spacetime in \cite{Firouzjahi:2023wbe, Firouzjahi:2024faf}.

The energy momentum tensor for the perturbations $\Phi(x^\mu)$ is given by
\ba
T^\mu_\nu = \partial^\mu \Phi \partial_\nu \Phi -\frac{1}{2} \delta^{\mu}_{ \nu}
\, \partial^\alpha \Phi \partial_\alpha \Phi \, .
\ea
In this section as our emphasis is on the quantum properties and the regularizations of the UV divergences, we consider only the energy density 
$\rho = -T^0_0$. The remaining components of the anisotropic energy density will be considered in details in next section where we study the backreactions induced by these quantum perturbations. 

Constructing the energy density, it is given by three contributions,
\ba
\rho= \rho_1 + \rho_2 + \rho_3 \, ,
\ea
in which 
\ba
\label{rho1-def}
\rho_1 \equiv \frac{1}{2 a(\tau)^2} \big(\partial_\tau \Phi(x^\mu) \big)^2 \, ,
\ea
\ba
\label{rho2-def}
\rho_2 \equiv \frac{1}{2 a(\tau)^2} \big(\partial_{\bf x} \Phi(x^\mu) \big)^2 \, ,
\ea
and
\ba
\label{rho3-def}
\rho_3 \equiv \frac{1}{2 r(\tau)^2} \gamma^{a b} \partial_a \Phi(x^\mu) \partial_b \Phi(x^\mu) 
 \, .
\ea
Before going into technicalities of calculating each of the above components, it is instructive to compare the magnitude of the above components near the singularity region, $r\rightarrow 0$.   Since both $\rho_1$ and $\rho_2$ deal with derivatives along the extended direction then on physical grounds we expect that $\rho_1 \sim \rho_2$. On the other hand, $\rho_3$ involves derivative along the compact directions so intuitively speaking 
$\rho_3 \sim (a^2/r^2) \rho_2$. Using the relations (\ref{ra-app}) we conclude that $\rho_3 \sim  a(\tau)^2 r(\tau)^{-2}  \rho_2 \sim (- \tau/GM)^{1/2} $. Correspondingly, we expect that near the singularity 
$\rho_3$ to be suppressed compared to 
$\rho_1$ or $\rho_2$ by a factor $ ( -\tau/GM)^{1/2} $.

We are interested in quantum expectation values such as $\langle \rho \rangle$. As is well known, there is no unique notion of the vacuum in a curved background such as in BH background where the spacetime contains an event horizon. In our analysis, we restrict ourselves to  the interior of the BH   with the mode function given in Eq. (\ref{Phi-pert}). The vacuum for the observer inside the BH is defined by $b_\bfk |0\rangle =0$. With our choice $C_2=0$, this vacuum is like the Bunch-Davies vacuum in inflationary setup.
Of course, since the coordinate $(r, \bar t)$ or equivalently $(\tau, x)$ do not cover the whole manifold, this vacuum is not well-defined globally. For example, one may instead use the vacuum associated to the Kruskal-Szekeres coordinate, i.e. the Hartle-Hawking vacuum  which is well-defined on both sides of the horizon. The difference in vacua employed by various observers is the prime reason for the  existence of Hawking radiation in BH background. In this work, we exclusively concentrate to the interior region with the vacuum $|0\rangle$ described above which is well-defined for all interior observers.

Let us start with $\langle \rho_2 \rangle$ which is easier to calculate. 
Also, to see the structure of divergences,   let us for the moment take $d=2$ (i.e. no dimensional regularization yet).
Using the mode function (\ref{Phi-pert}) and performing the summation  over $m$ as given in Eq. (\ref{sum-Ylm}), we obtain
\ba
\langle \rho_2(\tau) \rangle = \frac{1}{2 \pi r(\tau)^2 a(\tau)^2} 
\sum_{\ell=0}^\infty (2 \ell +1) 
\int_0^\infty \frac{d \omega}{ 2 \pi} \omega^2 | g_\omega(-\omega \tau)|^2 
\quad \quad (d=2) \, .
\ea
As usual in QFT analysis, the above expectation value suffers from the UV divergences. Since we decomposed the spatial directions into the form 
$R\times S^2$,  we have two different types of UV divergences. The first divergence is the infinite sum $\sum_{\ell=0}^\infty (2 \ell +1) $ associated to the azimuthal directions  while the other divergence is due to the integral of 
the high frequency mode $\omega$. Fortunately,  we are able to regularize both divergences as follows. 

The divergence associated to the infinite sum $\sum_{\ell=0}^\infty (2 \ell +1) $ can be dealt with using the zeta function regularization \cite{Hawking:1976ja, Elizalde:1994gf} scheme.
More specifically, we have 
\ba
\label{zeta1}
\sum_{\ell=0}^\infty (2 \ell +1) = -\zeta(-1) = \frac{1}{12} \, .
\ea

The UV divergence over the high frequency mode can be regulated via dimensional regularization as we advertised before.  For this, we replace 
\ba
\dd^{d-1} \bfk \rightarrow \omega^{d-2} \mu^{2-d}  \dd \omega \, \dd^{d-2} \Omega \, ,
\ea
in which $\dd^{d-2} \Omega$ represents the volume element of  a unit 
$(d-2)$-dimensional sphere with the volume
\ba
V_{d-2} \equiv \int \dd ^{d-2} \Omega = \frac{ 2 \pi^{ \frac{d-1}{2} } }{\Gamma(\frac{d-1}{2})}
\, ,
\ea
and  $d= 2 -\epsilon$ in which $\epsilon \rightarrow 0$ is the small parameter in dimensional regularization expansion.   The mass scale $\mu$ is introduced to keep track of the correct dimension of the physical quantities. 
Putting all together, we obtain 
\ba
\langle \rho_2(\tau) \rangle = \frac{-\zeta(-1) \mu^{2-d}}{2 \pi r(\tau)^2 a(\tau)^d} V_{d-2}
\int_0^\infty \frac{d \omega}{ (2 \pi)^{d-1}} \omega^d | g_\omega(-\omega \tau)|^2  \, .
\ea
The above integral can be performed analytically, yielding\footnote{We use the Maple computational software to calculate the integrals.}
\ba
\label{rho2-eq}
\langle \rho_2(\tau) \rangle = \frac{-\zeta(-1)\, (d-1)  \cos( \pi \nu) }{2^{d+2} \pi^{2+ \frac{d}{2}} a(\tau)^d r(\tau)^2 (-\tau)^d} \mu^{2-d} \Gamma\big( \frac{-d}{2} \big)  \Gamma\big(  \frac{5d+2}{8} \big)    \Gamma\big( \frac{3d+6}{8} \big) \, .
\ea
The divergence of the above expression appears in $\Gamma\big( \frac{-d}{2} \big)$ when $d\rightarrow 2$. Now, expanding to leading order in $d=2-\epsilon$ we obtain 
\ba
\langle \rho_2(\tau) \rangle = \frac{\zeta(-1) }{32 a(\tau)^2 r(\tau)^2 \pi^2 \tau^2} 
\Big( \frac{1}{\epsilon} + \ln\big( - \mu \tau a (\tau) \big) \Big) +  {\cal O}(\epsilon) \, ,
\ea
in which we have absorbed some numerical constants into the mass scale parameter $\mu$ in the final result. As expected, there is a pole at $\epsilon=0$ which highlights the UV divergence which should be regularized by appropriate counter terms as in standard QFT analysis. In addition, the appearance of the logarithmic term $\ln\big(-\mu \tau \big)$ is expected which is the hallmark of the dimensional regularization scheme. 

Discarding the divergence term via regularization, the regularized value of 
$\langle \rho_2(\tau) \rangle$ is obtained to be
\ba
\langle \rho_2(\tau) \rangle_{\mathrm{reg}} = 
\frac{\zeta(-1) }{32 a(\tau)^2 r(\tau)^2 \pi \tau^2} 
 \ln(-\mu \tau a(\tau)) \, .
\ea
The above expression is valid for all region inside the BH. We are mostly interested in the value of the energy momentum tensor near the singularity 
$\tau \rightarrow 0$ where the expectation values like $\langle \rho \rangle $
grow with negative power of $\tau$. Using near singularity relations (\ref{ra-app}), we obtain
\ba
\langle \rho_2(\tau) \rangle_{\mathrm{reg}} \equiv 
c_2 \ln(-\mu \tau) (-\tau)^{-\frac{5}{2}} \, ,
\ea
in which $c_2 \propto (G M)^{-\frac{3}{2}} $ is a constant involving $\zeta(-1)$  and other numerical factors. In addition, we have absorbed a factor 
$(G M)^{1/4}$ into $\mu$ which is an arbitrary scale at this level. 
We observe that  near the singularity $\langle \rho_2(\tau) \rangle_{\mathrm{reg}}$ grows like 
$(-\tau)^{-\frac{5}{2}} \sim r^{-5}$  with logarithmic running.
The factor $\ln(-\mu \tau) $  is the hallmark of the quantum corrections. 

Now we calculate $\langle \rho_1(\tau) \rangle$  which  from 
Eq. (\ref{rho1-def})  is given by
\ba
\langle \rho_1(\tau) \rangle = \frac{-\zeta(-1) \mu^{2-d}}{2 \pi a(\tau)^2} V_{d-2}
\int_0^\infty \frac{d \omega\,  \omega^{d-2}}{ (2 \pi)^{d-1}}  \Bigg| \frac{\dd  }{\dd \tau}
\Big(\frac{a(\tau)^{\frac{2-d}{2}}}{r(\tau)}g_\omega(-\omega \tau) \Big) \Bigg|^2  \, ,
\ea
in which we have used Eq. (\ref{zeta1}) to perform the sum over $\ell$ as before. Performing the integral using the dimensional regularization approach with the mode function given in Eq. (\ref{g*-rt}), and considering near the singularity region, 
we obtain
\ba
\label{rel1}
\langle \rho_1(\tau) \rangle = \frac{3(2+ d)}{6-5 d} 
\langle \rho_2(\tau) \rangle \, .
\ea
Regularizing $\langle \rho_1(\tau) \rangle$ by discarding the divergent $\epsilon^{-1}$ term we obtain 
\ba
\langle \rho_1(\tau) \rangle_{\mathrm{reg}} =  
c_1 \ln(-\mu \tau) (-\tau)^{-\frac{5}{2}} \, ,
\ea
in which $c_1$ is a numerical constant with the dimension $c_1 \sim c_2 \sim (G M)^{-\frac{3}{2}}$. Naively, Eq. (\ref{rel1}) may suggest that $c_1= - 3 c_2$ when $d=2$. However, this is misleading since we should actually set $d=2 -\epsilon $ which brings an additional contribution. More specifically, we have
\ba
\langle \rho_1(\tau) \rangle_{\mathrm{reg}}+ 3  \langle \rho_2(\tau) \rangle_{\mathrm{reg}} = \frac{3 \zeta(-1)}{128 \pi^2 (G M)^{\frac{3}{2}} (-\tau)^{\frac{5}{2} } } \, .
\ea


Now we calculate $\langle \rho_3 \rangle$. For this purpose, first note that 
$\gamma^{a b} \nabla_a \nabla_b \Phi = -\ell (\ell+1) \Phi$  in which $\gamma_{a b}$ is the metric on a unit two-sphere and $\{ a, b\}= \{ \theta, \phi\}$. This yields 
\ba
\langle \rho_3 \rangle &=& \frac{1}{2 r(\tau)^2} \gamma^{a b} \langle \partial_a \Phi \partial_b \Phi \rangle \nonumber\\
&=&   \frac{1}{2 \pi r(\tau)^4 a(\tau)^{d-2}} 
\sum_{\ell=0}^\infty (2 \ell +1) \ell (\ell+1) 
\int_0^\infty \frac{d \omega \omega^{d-2}}{ (2 \pi)^{d-1}} | g_\omega(-\omega \tau)|^2  \, .
\ea
There are a number of important differences compared to the case of $\langle \rho_2 \rangle$. First, we have a non-trivial sum over $\ell$ which requires more sophisticated form of $\zeta$ function regularization 
scheme \cite{Elizalde:1993ud, Elizalde:1994gf}. However, the important difference is that we have a weaker dependence 
on  $a(\tau)$ while the dependence on $r(\tau)$ is stronger. This is because 
$a(\tau)$ is the scale factor along the extended spatial direction while 
$r(\tau)$ represents the scale factor along the compact dimensions. 
Now, using the scalings of $a(\tau)$ and $r(\tau)$ near the singularity given in Eq. (\ref{ra-app}), we find that $\langle \rho_3 \rangle \propto (G M)^{-2} \tau^{-2}$. Upon performing the zeta regularization over the sum over $\ell$ and dimensional regularization over the frequency,  $\langle \rho_3(\tau) \rangle_{\mathrm{reg}}$ is obtained to have the following form,
\ba
\langle \rho_3(\tau) \rangle_{\mathrm{reg}} = c_3 \ln(-\mu \tau)\,  \tau^{-2} \, ,
\ea 
in which $c_3 \propto (G M)^{-2} $. 

Comparing the scaling of $\langle \rho_3 \rangle $  to $\langle \rho_1 \rangle $ or $\langle \rho_2 \rangle $ we observe that the $\tau$-dependence  in $\langle \rho_3 \rangle $ is milder compared to  latter by a factor 
$(-\tau/GM)^{1/2}$ as concluded previously based on qualitative arguments.  Therefore, near the singularity region we can neglect the contributions of $\langle \rho_3 \rangle $ compared to those of $\langle \rho_1 \rangle $
and $\langle \rho_2 \rangle $.

Armed with the results for $\langle \rho_i \rangle$ obtained above,  we can calculate their backreactions on the background metric. 


\section{Anisotropic $T^\mu_\nu$ and Backreactions}
\label{backreaction}

In the previous section we have calculated $\langle \rho \rangle_{\mathrm{reg}} $. 
To complete the study, we need to calculate other components of the energy momentum tensor. Since we are done with the regularization, we set $d=2$.

As the background is homogenous but anisotropic, the  energy momentum tensor induced by quantum fluctuations 
takes the form of  an imperfect fluid,  given by 
\cite{Ellis:1998ct} 
\ba
T_{\mu\nu} = (\rho + P) u_{\mu} u_{\nu} + P g_{\mu \nu} + q_{\mu} u_{\nu} +  q_{\nu} u_{\mu} + \pi_{\mu \nu} \, ,
\ea
with the additional conditions
\ba
u^{\mu}u_{\mu} =-1, \quad 
q_{\mu} u^{\mu}=0, \quad \pi^{\mu}_{\mu} =0 ,\quad \pi_{\mu \nu} = \pi_{\nu \mu} , \quad \pi_{\mu \nu} u^{\mu}=0 \, .
\ea
In the above decomposition, $\rho$ is the energy density, $P$ is  the isotropic pressure, $u^{\mu}$ is the fluid's comoving four velocity, $\pi_{\mu \nu}$ is the anisotropic pressure (stress) and $q^{\mu}$ is the heat conduction which also measures the energy flux relative to $u^{\mu}$. In our setup we expect  no heat flow for the fluid so $q_{\mu}=0$.

From the symmetry of the cosmological background, we obtain
\ba
u^{\mu} = (a(\tau), 0, 0, 0) \, .
\ea
In the absence of the heat flow, the energy momentum tensor takes the diagonal form 
\ba
T^{\mu}_{\nu} = \mathrm{diag} \left(   -\rho, P + \pi^{1}_{1}, P+ \pi^{2}_{2} , P+ \pi^{3}_{3}
\right) \, .
\ea
As the compact dimensions are spherically symmetric, we impose  $\pi^{2}_{2}= \pi^{3}_{3}$. Combined with the traceless condition, $\pi^{\mu}_{\mu}=0$, we obtain
\ba
\label{T-mu-nu}
T^{\mu}_{\nu} = \mathrm{diag} \left(   -\rho, P - 2 \Pi, P+\Pi , P+ \Pi \right) \, ,
\ea
where the anisotropic pressure is now controlled by a single scalar quantity, $\Pi\equiv \pi^{2}_{2}$. 

We are interested in vacuum expectation values of $\langle T^\mu_\nu \rangle$. From the above form of $T^\mu_\nu$ we need to calculate 
$\langle \rho \rangle$,  $\langle P \rangle$ and $\langle \Pi \rangle$.
The first quantity was already calculated in previous section 
as 
\ba
\label{rho-eq}
\langle \rho \rangle = 
\langle \rho_1 \rangle+ \langle \rho_2 \rangle+ \langle \rho_3 \rangle \, .
\ea  
As we shall see, $\langle P \rangle$ and $\langle \Pi \rangle$ can be expressed in terms of $\langle \rho_i \rangle$ as well. 

Let us consider the azimuthal components of $\langle T^\mu_\nu \rangle$.
From the spherical symmetry of the azimuthal directions we  have  $\langle \nabla_a \Phi \nabla_b \Phi \rangle = \kappa \gamma_{a b}$
in which $\kappa$ is a constant \cite{Firouzjahi:2022vij}. To find the value of $\kappa$, contract this relation by the inverse metric $\gamma^{ab}$, yielding  
\ba
\langle \nabla_a \Phi \nabla_b \Phi \rangle =  \gamma_{a b} \langle \rho_3 \rangle \, .
\ea
From this relation, we obtain
\ba
\label{Tab}
\langle T^a_b \rangle =  \delta^a_b \Big( \langle \rho_1 \rangle - 
\langle \rho_2 \rangle \Big) \, , \quad \quad \{ a, b\}= \{ \theta, \phi\} \, .
\ea
On the other hand, for $\langle T^x_x \rangle$ component we have
\ba
\label{Txx}
\langle T^x_x \rangle =   \langle \rho_1 \rangle + 
\langle \rho_2 \rangle  -  \langle \rho_3 \rangle \, .
\ea
Comparing Eqs. (\ref{Txx}) and (\ref{Tab}) with the general form of anisotropic 
$T^\mu_\nu$ given in Eq. (\ref{T-mu-nu}), we obtain
\ba
\label{P-eq}
\langle P\rangle =  \langle \rho_1 \rangle -\frac{1}{3} \langle \rho_2  \rangle 
- \frac{1}{3}\langle \rho_3 \rangle  \, ,
\ea
and
\ba
\label{Pi-eq}
\langle \Pi \rangle =   -\frac{2}{3} \langle \rho_2  \rangle 
+ \frac{1}{3}\langle \rho_3 \rangle   \, .
\ea

Our results for the values of  $\langle T^\mu_\nu \rangle$ were general so far. Now, we consider our limit of interests where $\tau\rightarrow 0$ with 
$\langle \rho_i \rangle_{\mathrm{reg}}$ as obtained in the previous section. 
 In this limit, the regularized components of the $\langle T^\mu_\nu \rangle$ 
 are given by 
\ba
\label{rho-sol}
\langle \rho \rangle_{\mathrm{reg}} = \Big[ (c_1+ c_2)(-\tau)^{\frac{-5}{2}} + c_3 \tau^{-2} \Big] \ln(-\mu \tau)\, ,
\ea
\ba
\label{P-sol}
\langle P \rangle_{\mathrm{reg}} =  \Big[ (c_1-\frac{ c_2}{3})(-\tau)^{\frac{-5}{2}} -\frac{c_3}{3}  \tau^{-2} \Big] \ln(-\mu \tau)\, ,
\ea
and
\ba
\label{Pi-sol}
\langle \Pi \rangle_{\mathrm{reg}} = \Big[ -\frac{2c_2}{3} (-\tau)^{\frac{-5}{2}}
+ \frac{c_3}{3} \tau^{-2}\Big] \ln(-\mu \tau) \, .
\ea
 
\subsection{Backreactions}

Having calculated the components of the  energy momentum tensor generated from the quantum perturbations, we can calculate the backreactions of this energy momentum tensor on the geometry.   

The Einstein field equations in the presence of the quantum  backreactions are given as usual by \cite{Weinberg:1972kfs}
\ba
R^\mu_\nu = 8 \pi G  \langle S^\mu_\nu \rangle_{\mathrm{reg}} \,  ,  \quad \quad S^\mu_\nu\equiv 
T^\mu_\nu -\frac{1}{2} \delta^\mu_\nu T \, .
\ea
In the absence of backreaction, $T^\mu_\nu=0$ so one deals with the vacuum Einstein field equations which of course yield the background Schwarzschild solution (\ref{metric-Sch}). In the presence of backreactions, we expect this solution to be modified. As we still  assume the spherical symmetry to hold in the presence of backreactions, it is convenient to use a metric ansatz  which takes the spherical symmetry into account. Motivated by the metric  (\ref{metric-Sch}) for the exterior of the BH, we propose  the following time-dependent spherical symmetric metric for the interior of the BH (see also \cite{Doran:2006dq, Gorji:2020ten} for a similar ansatz), 
\ba
\label{ansatz1}
ds^2 = -\frac{dt^2}{A(t)} + B(t) d x^2 + t^2 d \Omega^2 \, .
\ea 
In the above ansatz $t$ plays the roles of $r$ in the exterior region. In particular, in the absence of backreactions, the solution for the functions
$A(t)$ and $B(t)$ are
\ba
\label{AB-sol}
A(t) = B(t) = -1+ \frac{2 GM}{ t} \, ,  \quad \quad ( T^\mu_\nu=0) \, .
\ea
Of course, instead of the  metric ansatz (\ref{ansatz1}), we could have chosen a different cosmological ansatz, such as 
\ba
\label{ansatz2}
ds^2 = a_1(\tau)^2 (- d\tau^2 + dx^2) + a_2(\tau)^2 d \Omega^2 \, .
\ea
This metric is the extension of the metric (\ref{BH-metric}) to the case where the backreactions are included with two unknown scale factors $a_1(\tau)$ and $a_2(\tau)$. However, it turns out that the field equations are easier with the ansatz (\ref{ansatz1}) which we consider  in the following analysis. 

From the combinations  $2R^\theta_\theta - R^x_x- R^t_t$ and $R^x_x - R^t_t$ of the Einstein fields equation associated to the metric (\ref{ansatz1})
we obtain the following equations,
\ba
\label{Ein1}
\frac{1}{t^2} \Big( t \dot A + A +1 \Big) &=& 8 \pi G \Big( 2 \langle \Pi \rangle_{\mathrm{reg}} - 
\langle P\rangle_{\mathrm{reg}} \Big) , \\
\label{Ein2}
\frac{A}{t} \Big( \frac{\dot B}{B} - \frac{\dot A}{A} \Big) &=&
8 \pi G \Big(  \langle \rho \rangle_{\mathrm{reg}} + \langle P\rangle_{\mathrm{reg}} -  2 \langle \Pi \rangle_{\mathrm{reg}} \Big) \, .
\ea
One can easily check that in the absence of the backreaction the solutions of
the above equations are given by Eq. (\ref{AB-sol}) as expected. We can solve the above equations perturbatively as follows:
\ba
A(t)= \bar A{(t)}+ \delta A(t) \, , \quad \quad  B(t)= \bar B{(t)}+ \delta B(t) \, ,
\ea
in which $\bar A= \bar B$ are the background solutions given in Eq. (\ref{AB-sol}). Starting with Eq. (\ref{Ein1}),  we obtain $\delta A$ as follows,
\ba
\label{A1-sol}
\delta A= \frac{8 \pi G}{t} \int^t d t' t'^2 \Big[  2 \langle \Pi(t') \rangle_{\mathrm{reg}} -  \langle P(t') \rangle_{\mathrm{reg}} \Big] \, .
\ea
Now using this solution into Eq. (\ref{Ein2}) we obtain 
\ba
\label{B1-sol}
\delta B= \delta A
+  8 \pi G \bar A(t)  \int^t d t' \frac{t'}{\bar A(t')} 
\Big[    \langle \rho(t') \rangle_{\mathrm{reg}}+  \langle P(t') \rangle_{\mathrm{reg}} -  2\langle \Pi(t') \rangle_{\mathrm{reg}} \Big] \, .
\ea
The above solutions for $\delta A$ and $\delta B$ were general with no assumption on the source terms in the right hand sides of Eqs. (\ref{A1-sol})
and (\ref{B1-sol}). Now, we consider our case of interests with $ \langle \rho \rangle_{\mathrm{reg}},  \langle P \rangle_{\mathrm{reg}}$ and $ \langle \Pi \rangle_{\mathrm{reg}}$ given in Eqs. (\ref{rho-sol}), (\ref{P-sol}) and
(\ref{Pi-sol}). To use these expressions for the source terms, we need to express $\tau$ in terms of $t$. From Eq. (\ref{ra-app}) we note that 
$\tau \simeq  -t^2/4 GM$ (note that in our treatment of the interior of the BH, $t$ plays the same roles as the coordinate $r$ outside the BH).
Correspondingly, the source terms are obtained to be  
\ba
\label{source1}
2\langle \Pi \rangle_{\mathrm{reg}} -  \langle P \rangle_{\mathrm{reg}}
=\Big[  -( c_1+ c_2) (-\tau)^{-5/2} + c_3 \tau^{-2} \Big] \ln(-\mu\,  \tau)
\simeq \alpha \ln(\mu\,  t)\,  t^{-5} \, ,
\ea
in which $\alpha \propto (c_1+ c_2) \sim G M$ is a constant and we have rescaled the undetermined scale $\mu$ as well.  
We have neglected the subleading term 
containing $c_3$ as long as $c_1 \neq -c_2$ which is the case in our analysis.  Similarly, for the other source term, we obtain
\ba
\langle \rho \rangle_{\mathrm{reg}}+   \langle P \rangle_{\mathrm{reg}} -2\langle \Pi \rangle_{\mathrm{reg}} = 2 ( c_1+ c_2) \ln(-\mu\,  \tau) (-\tau)^{-5/2}=  -2 \alpha  \ln(\mu\,  t)\,  t^{-5} \, .
\ea
With the above source terms at hand, we can calculate $\delta A$ and $\delta B$. Considering the near singularity limit $t \rightarrow 0$, from Eq. (\ref{A1-sol}) we obtain, 
\ba
\delta A \simeq \frac{8 \pi G \alpha }{t}  \int^t dt' \, \ln(\mu\,  t')  t'^{-3} \simeq -\frac{4 \pi G \alpha}{t^3} \ln(\mu t)  \, ,
\ea
in which we have absorbed some numerical constant in the scale $\mu$ as well. 
Using this solution in  equation (\ref{B1-sol}), we obtain
\ba
\delta B \simeq \delta A - \frac{16 \pi G \alpha}{t} \int^t dt' t'^{-3} \simeq 
\frac{4 \pi G \alpha}{t^3} \ln(\mu t) \, .
\ea 

In conclusion, the metric functions to leading order in backreaction effects are given by
\ba
\label{back-A}
A(t) &\simeq& -1+ \frac{2 G M}{t} -\frac{4 \pi G \alpha}{t^3} \ln(\mu t)  \, , \\
\label{back-B}
B(t) &\simeq& -1+ \frac{2 G M}{t} +\frac{4 \pi G \alpha}{t^3} \ln(\mu t) \, .
\ea
with $\alpha \propto GM$. As mentioned before, the appearance of the logarithmic correction $\ln(\mu t) $ is reassuring, highlighting  the nature of the regularized quantum corrections. 

\subsection{Physical Implications}

Having obtained the backreaction in the metric, one may ask for some physical implications. 

One question to ask is  what is the threshold where the backreaction contributions become comparable to the background solution, $\delta A \sim \bar A$?  For this purpose, let us consider the metric along the azimuthal direction with the line element $t^2 d \Omega^2$.  
This may define the shortest distance (radius  of two-sphere  near the singularity) that we can trust our approximate solution before the backreactions overcome the background solution. Otherwise, this can define the scale in which  new physics may enter to resolve the singularity. Defining this radius (actually the time scale towards singularity) by $t_c$, it is given by
\ba
\frac{2 G M}{t_c} \sim \frac{4 \pi G \alpha}{t_c^3} \ln(\mu t_c)
\quad \rightarrow  \quad
 t_c  \sim 
\big( \frac{\alpha}{M}\big)^{\frac{1}{2}} \, , 
\ea
in which in obtaining the final estimation, we have neglected the subleading logarithmic correction involving the scale $\mu$.  With $\alpha \sim GM$, this timescale is obtained to be $t_c \sim\sqrt G \sim 1/M_P$, i.e. the Planck time. This is somewhat expected as  
$M_P$ defines the scale of quantum gravity. Of course the above estimation is rough as we have considered only  near the singularity limit of our approximate solution. Treating the analysis more carefully, one may obtain a somewhat larger scale for $t_c$ prior to the singularity when  the backreaction becomes important. This requires a careful investigation which is beyond the scope of the current analysis.

The other interesting question is what the exterior observer interpret the corrections in the metric? Based on our starting discussion, for an exterior observer the roles of the timelike and spacelike coordinates are switched. This suggests the speculation 
 that for the exterior observer in which $t\rightarrow r$ and $x\rightarrow \bar t$, the line element with the quantum corrections is given by
\ba
\label{metric-ext}
ds^2 = -{\cal B}(r) \dd {\bar t\, }^2 + \frac{\dd r^2}{{\cal A}(r)} + r^2 d \Omega^2 \, \quad  \quad \quad  (r> 2 G M) \, ,
\ea
with the functions ${\cal A}(r)$ and ${\cal B} (r)$ are given  as follows
\ba
\label{back-A2}
{\cal A}(r) =-A(r) &\simeq& 1- \frac{2 G M}{r} +\frac{4 \pi G \alpha}{r^3} \ln(\mu r)  \, , \\
\label{back-B2}
{\cal B}(r) = -B(r) &\simeq& 1- \frac{2 G M}{r} -\frac{4 \pi G \alpha}{r^3} \ln(\mu r) \, .
\ea
Interestingly, the above speculative results are in agreement with the results for the  quantum backreactions for the exterior region \cite{Duff:1974ud, Bjerrum-Bohr:2002fji,  Donoghue:2001qc, Calmet:2017qqa}
who obtained   $1/r^3$ profile with possible logarithmic running 
for the metric components. This similarity is encouraging,  indicating that the effects of backreactions from the quantum fields in the interior of the BH are genuine. 

An interesting corollary from the above result is that the quantum corrections  in the interior of BH can be probed by the exterior observer. Indeed, the exterior observer equipped with the quantum modified metric (\ref{metric-ext}) can observationally distinguish this metric from the classical Schwarzschild metric. This is intriguing, implying that  the effects occurring inside the event horizon is accessible to the exterior observer. Of course, like in Hawking radiation, this is a pure quantum effect.  Having said this, the quantum corrections in Eqs. (\ref{back-A2}) and (\ref{back-B2}) are so small that 
they seem impossible to be probed observationally. 

With the quantum corrections in background metric included,  it is insightful to look at various of its physical effects. First, the quantum corrections to the metric will modify the position of the event horizon. The new position of the event horizon $R_S'$ from the solution of 
${\cal B}(r)=0$  is given by 
\ba
R_S'\simeq R_S - \frac{\pi \alpha}{G M^2}   \ln( 2 \mu G M) \, . 
\ea
With $\alpha \sim GM$ the fractional corrections in the position of the event horizon is at the order  $(M R_S)^{-1} \sim (M_P/M)^2$ which is extremely small and practically  unobservable for astrophysical BHs. Since the fractional correction induced by parameter $\alpha$ is tiny, the quantum corrections can not modify the number of event horizons so there is only  one event horizon. 

The surface gravity associated to the metric (\ref{metric-ext}) is given by 
\cite{Carroll:2004st, Visser:1992qh}
\ba
\kappa = \frac{{\cal B}'}{2}\sqrt{\frac{ {\cal A}}{ {\cal B}}}\, \Big|_{R_S} \, ,
\ea
which to leading order in $\alpha$,  is obtained to be
\ba
\kappa\simeq \frac{1}{4 G M}+ \frac{\pi \alpha}{8 G^3 M^4} \big(3 \ln(2 \mu G M) - 1 \big)\, .
\ea
Correspondingly, the fractional corrections in surface gravity is at the 
order $\alpha/G^2 M^3 \sim (M_P/M)^2$ which is extremely small for astrophysical BHs. Related to this issue, note that since the fractional corrections in event horizon and surface gravity is very small, we can not hope to have a extremal black hole with a vanishing surface gravity. Physically, this is because the parameter $\alpha$ is not independent of $M$ as $\alpha \sim GM$. Therefore, we can not expect to trade $M$ with 
$\alpha$ as an independent parameter  to obtain an extremal black hole. 

We can calculate the ADM mass associated to the metric (\ref{metric-ext}) as well. Since the quantum correction falls off like $\ln(\mu r)/r^3$, it is expected that its contribution in surface integrals at infinity in calculating the ADM mass to vanish.  Indeed, performing the analysis \cite{Poisson:2009pwt}, we confirmed that the ADM mass is simply $M$.

Another question is whether the quantum corrections can resolve the singularity at the center of BH.  Since we have treated the quantum effects as perturbative backreactions, we do not expect that they can resolve the singularity at this stage. Indeed, calculating the Ricci scalar associated to the interior metric (\ref{ansatz1}) with the metric functions (\ref{back-A}) and (\ref{back-B}) for $t \rightarrow 0$ we obtain
\ba
R \simeq 4 \pi G \alpha \big( 3- 2 \ln( \mu t) \big) t^{-5} \, .
\ea
Similarly, calculating the Kretschmann scalar near the center, we obtain
\ba
R_{\mu \nu \kappa \rho } R^{\mu \nu \kappa \rho } \simeq 
16 \pi^2 G^2 \alpha^2 \big( 53- 192 \ln(\mu t) + 184 \ln( \mu t)^2 \big) t^{-10}\, .
\ea
Both of the above scalar invariants diverge for $t\rightarrow 0$ so the perturbative quantum corrections do not resolve the singularity at the center of the BH. 

Before closing this section, we comments that the metric (\ref{metric-ext}) is obtained in the semi-classical approximation in which the matter  field is treated quantum mechanically while the geometry is still at the classical level. In this view, this solution can not capture the whole quantum gravity effects. However, this result may be useful  towards the understanding of a full quantum gravity solution. It is speculated that the final remnant of the BH after the completion of Hawking radiation is a quantum primordial BH with the mass around $M_P$. This can have interesting observational implications, for example as the source of dark matter \cite{MacGibbon:1987my, Domenech:2023mqk}.  In addition, as we discussed previously,  the backreactions by the quantum field modify the position of the event horizon and the BH surface area. This in turn induces corrections in BH entropy and its thermodynamics properties \cite{Bonanno:2000ep, Feng:2015jlj, Feng:2016tki, Xiao:2021zly}.However, as we have seen, the fractional correction is physical quantities such as the surface gravity and the position of the event horizon scales like $\alpha/G^2 M^3 \sim (M_P/M)^2$  which is extremely small for astrophysical BHs. Therefore, we expect that the corrections in BH thermodynamics are small and unobservable for astrophysical BHs.

\section{Summary and Discussions}

In this work we have studied the propagation of quantum field perturbations in the interior of a Schwarzschild black hole. As reviewed in section 
\ref{BH-cosmology}, the interior of the BH looks like an anisotropic cosmological background with the structure of $R^2 \times S^2$. As time moves forward, the extended direction with the scale factor $a(\tau)$ is expanding while the compact direction, with the scale factor $r(\tau)$, is shrinking. Eventually, we encounter the big crunch singularity when 
$r(\tau) \sim (-G M\tau)^{1/2}   \rightarrow 0 $. 

In our treatment, the quantum perturbations are initially generated near the horizon $\tau \rightarrow - \infty$ with the positive frequency solution $e^{-i \omega \tau}$. In the interior region, this plane wave encounters the effective potential $V_{\mathrm{eff}}$ which has a non-trivial time-dependent profile. We are able to find the analytic solution for the high frequency mode where the effects of the term $\ell (\ell+1) $ can be neglected. In this limit the solution is given in terms of the Hankel functions $H_\nu^{(1)}$ and
$H^{(2)}_\nu$ of which only the former can match to the incoming positive frequency mode  near the horizon. With the profile of the mode function at hand, we have calculated the vacuum expectation values of the components of the anisotropic energy momentum tensor such as $\langle \rho \rangle$, 
$\langle P \rangle$ and $\langle \Pi \rangle$. To regularize the UV divergences, we have employed the $\zeta$ function regularization for the infinite sum  over $\ell$ and the dimensional regularization scheme for the integral 
over ${\bf k}$. Upon performing the regularizations, we have found that  
near the big crunch singularity  $\langle \rho \rangle_{\mathrm{reg}} \sim \langle P \rangle_{\mathrm{reg}} \sim \langle \Pi \rangle_{\mathrm{reg}} \sim (G M)^{-3/2} (-\tau)^{\frac{-5}{2}} \ln(-\mu \tau)$. The appearance of the logarithmic term $ \ln(-\mu \tau)$ is reassuring, indicating the quantum nature of the corrections. 

Having obtained $ \langle T^\mu_\nu\rangle_{\mathrm{reg}}$, we have calculated the backreactions of the quantum perturbations on the background metric. We have found the interesting result that  
$\delta A = -\delta B \sim G^2 M \ln(\mu t) /t^3$. To trust our perturbative treatment, we require that the backreactions on the background solution to be small, requiring that $\delta A \ll \bar A$. This in turn yields the time scale (or the radius of the two-sphere)  $t_c \sim \sqrt{G}$
beyond which the backreactions become strong. It is possible that the backreactions become important sooner (i.e. at larger radius) than what we obtained here since we have imposed various simplification assumptions to solve the system analytically. Alternatively, one may argue that the quantum corrections, like what are calculated here, may play important roles  to resolve the singularity. Whether or not these quantum corrections can resolve the classical singularity require more sophisticated QFT analysis with higher curvature terms included in the gravitational action as well.

One intriguing implication of our result is that the corrections induced on the metric by the quantum effects in the interior of the BH are, in principle, accessible to the exterior observer. 
This is somewhat counterintuitive as  it is   usually believed that events occurring in the interior of the BHs are hidden to the exterior observer. However, this picture seems to be valid only classically and in the presence of quantum effects, as in the case of Hawking radiation, one may have to revise the classical interpretation about the roles of the event horizon. Having said this, the fractional correction in the metric and in the position of the horizon is 
at the order $(M_P/M)^2$ which is extremely small for astrophysical BHs.

There are a number of  directions in which the current analysis can be extended.  In the current analysis we have considered the quantum perturbations of a real scalar field in the interior of the BH. However, this analysis can be extended to other fields such as the $U(1)$ gauge (vector) fields or tensor perturbations as well. As in  classical quasi-normal mode analysis, the effective potential $V_{\mathrm{eff}}$ will now depend on the spin of the field $s$.   We would like to come back to the study of quantum perturbations for the vector ( $s=1$) and tensor perturbations ($s=2$)  in future.  Finally, we have calculated the vacuum expectation values with respect to the vacuum $| 0 \rangle$ defined for the interior observer. On the other hand, the vacuum associated to the Kruskal observer $| 0\rangle_K $ (or the Hartle-Hawking vacuum)  has the advantage that it is well-defined for both interior and exterior observers. These two vacua are related to each other via the Bogolyubov transformations. 
It is therefore an interesting question to examine what the Kruskal observer,
equipped with the Hartle-Hawking vacuum, measures for  the expectation values.  

\vspace{0.7cm}

{\bf Acknowledgments:}  We thank Mohammad Ali Gorji, Raihaneh Moti, 
Haidar Sheikhahmadi and Alireza Talebian  for useful discussions and correspondences. 
This work is supported by the INSF  of Iran under the grant  number 4022911.


\bibliography{BH-interior-PRD2}{}

\bibliographystyle{JHEPNoTitle}

\end{document}